\documentstyle[preprint,aps,epsfig]{revtex}
\tightenlines
\begin{document}
\title{ $ B_{c} $ meson and the light-heavy quarkonium spectrum }
\author{ M. Baldicchi and G.M. Prosperi }
\address{
Dipartimento di Fisica dell'Universit\`{a} di Milano, \\
and I.N.F.N., Sezione di Milano, Italy \\ }
\date{\today}
\maketitle
\begin{abstract}
We compute the $ c \bar{b} $ spectrum
from a first principle Salpeter equation
obtained in a preceding paper.
For comparison we report also
the heavy-light quarkonium spectrum
and the hyperfine separations previously
presented only in a graphical form.
Notice that all results are parameter free.
\end{abstract}
\setcounter{equation}{0}
\section*{}
The spectrum and properties of the $ c {\bar b} $ systems have
been calculated various times in the past in the framework of
the heavy quarkonium theory \cite{vecchi}.
However the recent
experimental observation of the $ B_{c}^{+} $ meson
\cite{bc1} has arisen new 
theoretical interest on the problem
\cite{bc2,fulcher,ukqcd}.
The mentioned spectrum has been considered again
either from the potential and the lattice simulation point of
view. A particular interesting quantity should be
the hyperfine splitting that as for the $ c {\bar c} $
case seams to be sensible to
relativistic and subleading corrections in $ \alpha_{\rm s} $.

For the above reasons it seems to us
worthwhile to present in this paper a calculation
of the $ c {\bar b} $ spectrum based on an effective
mass operator with full relativistic kinematics which we have
obtained in previous works and applied with a certain success to
a fit of the entire quarkonium spectrum, heavy-heavy,
light-light and light-heavy cases with the exception however
of the $ c {\bar b} $ case \cite{ioprosp1,ioprosp2}.
For comparison and completeness we report also numerical results
for the light-heavy spectrum which we have given previously
only in a graphical form.

The mass operator was obtained by a three dimensional reduction of the 
$ q \bar{q} $ Bethe-Salpeter equation introduced in \cite{prosp96}.
It has the quadratic form
$ M^{2} = M_{0}^{2} + U $, with a kinetic part
$ M_{0} = w_{1} + w_{2} = \sqrt{ m_{1}^{2} + {\bf k}^{2} }
+ \sqrt{ m_{2}^{2} + {\bf k}^{2} } $
and a ``potential'' that in terms of the istantaneous approximation of 
the B-S kernel is given by
\begin{equation}
\langle {\bf k} | U | {\bf k}^{\prime} \rangle =
\frac{1}{( 2 \pi)^{3} } \;
\sqrt{ \frac{ w_{1} + w_{2} }{ 2 \, w_{1} w_{2} } } \;
\hat{I}_{\rm inst } ( {\bf k} , {\bf k}^{\prime} ) \;
\sqrt{ \frac{ w_{1}^{\prime} + w_{2}^{\prime} }{
2 \, w_{1}^{\prime} w_{2}^{\prime} } },
\label{potu}
\end{equation}
$ {\bf k} $ denoting the momentum of the quark in the
centre of mass frame,
and $ i = 1,2 $ the quark and the antiquark. 

The B-S equation was derived from QCD first principles, taking 
advantage of the Feynman-Schwinger path integral
representation for the ``second order'' quark propagator in an
external field \footnote{Second order propagator in the sense
that it is defined by a second order differential equation;
the quadratic form of the mass operator derives essentially
from this fact.}.
The only assumption used consisted in writing the logarithm
of the Wilson loop correlator 
$ W = \frac{1}{3} \left\langle {\rm Tr} \, P \exp \!
\left( \oint \! dx^{ \mu } A_{ \mu } \right) \right\rangle $,
as the sum of its perturbative expression and an area term
\begin{equation}
  i \ln W = i ( \ln W )_{ \rm{pert} } + \sigma S_{ \rm{min} },
\label{wilson}
\end{equation}
$\sigma$ denoting the string tension.

An explicit expression for $ U $ is given in Ref. \cite{ioprosp2}.
The perturbative part of such quantity was evaluated at the
lowest order in $\alpha_{\rm s}$. However for $\alpha_{\rm s}$
we have used the standard
running expression
\begin{equation}
\alpha_{\rm s} ( {\bf Q} ) = \frac{4 \pi}{ ( 11 -
\frac{2}{3} N_{\rm f} )
\ln \frac{ {\bf Q}^{2} }{ \Lambda^{2} } }
\label{corre}
\end{equation}
(with $ N_{\rm f} = 4 $ and $ \Lambda = 200 \; {\rm MeV} $)
cut at a maximum value $ \alpha_{\rm s} ( 0 ) $, to treat
properly the infrared region \cite{marchesini}. This amount
to include important
perturbative subleading contributions.

Notice that, contrary to all the usual potential models,
we have given the
light quark current and not component masses in our
treatment. Component masses of the usual order of magnitude
can be recovered at a successive step
as effective values in a semirelativistic reformulation
\cite{ioprosp1}.
Actually we have fixed such masses on typical values,
$ m_{u} = m_{d} = 10 \; {\rm MeV} $, $ m_{s} = 200 \; {\rm MeV} $,
which are not adjusted in the fit (the results depend essentially on
$ \langle k \rangle $ and are very little affected by the precise
value of the light quark masses).
The other parameters of the theory are assumed as:
$ m_{c} = 1.394 \; {\rm GeV} $, $ m_{b} = 4.763 \; {\rm GeV} $,
$ \sigma = 0.2 \; {\rm GeV}^{2} $, $ \alpha_{\rm s} ( 0 ) = 0.35 $.
The first two are chosen in order to reproduce correctly the
$ J/\Psi $ and the $ \Upsilon ( 1 S ) $
masses, the string tension to give
the correct slope for the Regge $ \rho $ trajectory,
$ \alpha_{\rm s} ( 0 ) = 0.35 $ to give the right
$ J/\Psi - \eta_c $ splitting.
Notice that, consequently, the results
reported in this paper are completely parameter free,
with the exception of the
$ c {\bar c} ( 1 S ) $ hyperfine splitting.

We have used in our calculations also the more conventional
``linear mass'' operator (or center of mass relativistic
Hamiltonian) $ M = M_{0} + V $ (where $ V $ is defined by
$ U = M_{0} V + V M_{0} + V^{2} $)
which makes easier a comparison with the usual phenomenological
models. If we neglect the $ V^{2} $ term, $ V $ is obtained
from Eq.(\ref{potu}) simply by the kinematical replacement
\begin{equation}
\sqrt{ \frac{ w_{1} + w_{2} }{ 2 \, w_{1} w_{2} } } \;
\sqrt{ \frac{ w_{1}^{\prime} + w_{2}^{\prime} }{
2 \, w_{1}^{\prime} w_{2}^{\prime} } } \rightarrow
\frac{1}{ 4 \, \sqrt{ w_{1} w_{2} w_{1}^{\prime} w_{2}^{\prime} } }.
\end{equation}
This is the form we have used in Ref. \cite{ioprosp1}
(for some state however
$ \langle V^{2} \rangle $ is not negligible). In the
calculations based on this linear formalism we have used
the same values for the light quark masses as before,
a fixed coupling constant
$ \alpha_{\rm s} = 0.363 $
and taken
$ m_{c} = 1.40 \; {\rm GeV} $,
$ m_{b} = 4.81 \; {\rm GeV} $
and
$ \sigma = 0.175 \; {\rm GeV}^{2} $.

Details on the numerical treatment of the eigenvalue 
equation are given in \cite{ioprosp1} and \cite{ioprosp2}.

In table \ref{tabbc} we have reported the $ c {\bar b} $
spectrum as obtained by the quadratic and the linear formalism,
together with the values
presented in Refs. \cite{fulcher} and \cite{ukqcd}.
The observed mass $ M(B_{c}) = 6.40 \pm 0.39 \pm 0.13 $ GeV
has to be referred to the $ 1 \, {^{1} {\rm S}_{0}} $
state. For such state
all calculations give very close results
and reproduce equally well the experimental
value within the errors.
Larger discrepancies among the various methods
occur for the excited states.

In table \ref{tablp} we have reported the spectrum
for light-heavy mesons obtained by our formalism in
numerical form.
We have considered the hyperfine structure but omitted
the fine one. We have also reported the quantity
$ \Delta_{\rm avg} $ defined as the average of the
deviations of the theoretical values from the experimental
data diminished by the experimental errors.
Obviously $ \Delta_{\rm avg} $ provides a
measure of the accuracy in reproducing
the data and give an idea of the precision one can
expect in the $ c {\bar b} $ case.

In table \ref{iperf}, finally, we have reported the hyperfine
splitting for the $ 1 S $ and $ 2 S $ states as obtained
in the quadratic formalism
and the $ \Delta_{\rm avg} $ quantity
even for the channels for which we do not reproduce
the results in full here.

Notice the strong discrepancies with the data in the
hyperfine splittings of the $ 1 S $
light-light cases. This is obviously due to the chiral symmetry
breaking problem and the related inadequacy of replacing the
quark full propagator in the B-S equation with the free form,
as implied in the three-dimensional reduction.
For the rest, the agreement is good for the states involving
light and $ c $ quarks, while the theoretical value tends
to be too large for states involving $ b $ quarks.

For comparison we can mention that in the linear formalism the
hyperfine splitting turns out less good, being e.g. 97 MeV for
$ c \bar{c} ( 1 S ) $, 111 MeV  for $ u \bar{c} ( 1 S ) $,
108 MeV  for $ c \bar{s} ( 1 S ) $.
Likely such difference has to be ascribed to relativistic and
$ \alpha_{\rm s} $ subleading effects, taken into account in
the quadratic formalism via Eq.(\ref{corre}).

In conclusion let us mention explicitly that
$ \Delta_{\rm avg} $ as reported in table \ref{iperf}
do not include the states $ c \bar{c} ( 4 S ) $
and $ b \bar{b} ( 6 S ) $, (which are largely above
threshold) and the
$ 1 \, {^{1} {\rm S}_{0}} $ and
$ 1 \, {^{1} {\rm P}_{1}} $
light-light states for the reasons recalled above.
\begin{table}
\centering
\caption{ $ b \bar{c} $ quarkonium systems. Experimental
$ B_{c} $ mass equal to 6.40 $ \pm $ 0.39 $ \pm $ 0.13 GeV. }
\begin{tabular}{ccccc}
States & quadratic formalism & linear formalism & Fulcher & Lattice \\
       & (GeV) & (GeV) &    (GeV)   &      (GeV)     \\
\hline
$ 1 \, {^{1} {\rm S}_{0}} $ & 6.258 & 6.293 & 6.286 &
$ 6.280 \pm 0.200 $ \\
$ 1 \, {^{3} {\rm S}_{1}} $ & 6.334 & 6.355 & 6.341 &
$ 6.321 \pm 0.200 $ \\
$ 2 \, {^{1} {\rm S}_{0}} $ & 6.841 & 6.848 & 6.882 &
$ 6.960 \pm 0.080 $ \\
$ 2 \, {^{3} {\rm S}_{1}} $ & 6.883 & 6.881 & 6.914 &
$ 6.990 \pm 0.080 $ \\
$ 3 \, {^{1} {\rm S}_{0}} $ & 7.222 & 7.221 & &  \\
$ 3 \, {^{3} {\rm S}_{1}} $ & 7.254 & 7.245 & &  \\
\hline
$ 1 \, {\rm P} $ & 6.772 & 6.762 & 6.754 &
$ 6.764 \pm 0.030 $ \\
$ 2 \, {\rm P} $ & 7.154 & 7.138 & &  \\
\hline
$ 1 \, {\rm D} $ & 7.043 & 7.025 & 7.028 &  \\
$ 2 \, {\rm D} $ & 7.367 & 7.346 & &  \\
\end{tabular}
\label{tabbc}
\end{table}
\begin{table}
\centering
\caption{ Light-Heavy quarkonium systems.}
\begin{tabular}{ccccc}
 States &  & experimental values &
 linear formalism & quadratic formalism \\
    &   &  (MeV)  &  (MeV)  &  (MeV)   \\
\hline
$ u \bar{c} $ &  &  &  &  \\
$ 1 \, {^{1} {\rm S}_{0}} $ &
$
\left\{
\begin{array}{c}
D^{\pm} \\
D^{0}
\end{array}
\right.
$
&
$
\left.
\begin{array}{c}
1869.3 \pm 0.5  \\
1864.5 \pm 0.5
\end{array}
\right\}
$
& 1890 & 1875  \\
$ 1 \, {^{3} {\rm S}_{1}} $ &
$
\left\{
\begin{array}{c}
D^{\ast}(2010)^{\pm} \\
D^{\ast}(2007)^{0}
\end{array}
\right.
$
&
$
\left.
\begin{array}{c}
2010.0 \pm 0.5  \\
2006.7 \pm 0.5
\end{array}
\right\}
$
& 2001 & 2020  \\
$ 2 \, {^{1} {\rm S}_{0}} $ & $ D^{\prime} $ &
2580 & 2556 & 2525  \\
$ 2 \, {^{3} {\rm S}_{1}} $ & $ D^{\ast \prime} $ &
2637 $ \pm $ 8 & 2615 & 2606  \\
\hline
$
1 \, {\rm P}
$ &
$
\begin{array}{c}
\left\{
  \begin{array}{c}
    D_{2}^{\ast} (2460)^{\pm}  \\
    D_{2}^{\ast} (2460)^{0}
  \end{array}
\right.
\\
\left\{
  \begin{array}{c}
    D_{1} (2420)^{\pm}  \\
    D_{1} (2420)^{0}
  \end{array}
\right.
\end{array}
$ &
$
\left.
\begin{array}{c}
  \begin{array}{c}
    2459 \pm 4   \\
    2458.9 \pm 2.0
  \end{array}
\\
  \begin{array}{c}
    2427 \pm 5   \\
    2422.2 \pm 1.8
  \end{array}
\end{array}
\right\}
$ & 2442 & 2475 \\
\hline
$ \Delta_{\rm avg} $ &  &  & 12 & 21 \\
\hline
\hline
$ u \bar{b} $ &  &  &  &  \\
$ 1 \, {^{1} {\rm S}_{0}} $ &
$
\left\{
\begin{array}{c}
B^{\pm} \\
B^{0}
\end{array}
\right.
$
&
$
\left.
\begin{array}{c}
5278.9 \pm 1.8  \\
5279.2 \pm 1.8
\end{array}
\right\}
$
& 5282 & 5273  \\
$ 1 \, {^{3} {\rm S}_{1}} $ &
$ B^{\ast} $ & 5324.8 $ \pm $ 1.8 & 5341 & 5339 \\
$ 2 \, {^{1} {\rm S}_{0}} $ &  &  & 5878 & 5893 \\
$ 2 \, {^{3} {\rm S}_{1}} $ &
$ B^{\ast \prime} $ & 5906 $ \pm $ 14 & 5916 & 5933 \\
\hline
$ 1 \, {\rm P} $ &  & 5825 $ \pm $ 14 & 5777 & 5792 \\
\hline
$ \Delta_{\rm avg} $ &  &  & 34 & 19 \\
\hline
\hline
$ s \bar{c} $ &  &  &  &  \\
$ 1 \, {^{1} {\rm S}_{0}} $ &
$ D_{s}^{\pm} $ & 1968.5 $ \pm $ 0.6 & 1999 & 1982 \\
$ 1 \, {^{3} {\rm S}_{1}} $ &
$ D_{s}^{\ast \pm} $ & 2112.4 $ \pm $ 0.7 & 2107 & 2120 \\
$ 2 \, {^{1} {\rm S}_{0}} $ &  &  & 2667 & 2617 \\
$ 2 \, {^{3} {\rm S}_{1}} $ &  &  & 2729 & 2698 \\
\hline
$
1 \, {\rm P}
$ &
$
\begin{array}{c}
D_{sJ} (2573)^{\pm} \\
D_{s1} (2536)^{\pm}
\end{array}
$ &
$
\left.
\begin{array}{c}
2573.5 \pm 1.7 \\
2535.35 \pm 0.34
\end{array}
\right\}
$ & 2528 & 2548 \\
\hline
$ \Delta_{\rm avg} $ &  &  & 21 & 9 \\
\hline
\hline
$ s \bar{b} $ &  &  &  &  \\
$ 1 \, {^{1} {\rm S}_{0}} $ &
$ B_{s}^{0} $ & 5369.3 $ \pm $ 2.0 & 5373 & 5364 \\
$ 1 \, {^{3} {\rm S}_{1}} $ &
$ B_{s}^{\ast} $ & 5416.3 $ \pm $ 3.3 & 5433 & 5429 \\
$ 2 \, {^{1} {\rm S}_{0}} $ &  &  & 5974 & 5985 \\
$ 2 \, {^{1} {\rm S}_{0}} $ &  &  & 6014 & 6024 \\
\hline
$ 1 \, {\rm P} $ &
$ B_{sJ}^{\ast} (5850) $ & 5853 $ \pm $ 15 & 5848 & 5859 \\
\hline
$ \Delta_{\rm avg} $ &  &  & 5 & 4 \\
\end{tabular}
\label{tablp}
\end{table}
\begin{table}
\centering
\caption{Hyperfine splitting (MeV)}
\begin{tabular}{ccccccccccc}
  & $ u \bar{c} $ & $ s \bar{c} $ & $ u \bar{b} $ & $ s \bar{b} $ &
$ c \bar{c} $ & ~ $ c \bar{b} $ ~
& ~ $ b \bar{b} $ ~ & $ u \bar{u} $ &
$ u \bar{s} $ & $ s \bar{s} $ \\
\hline
$ 1 \, {\rm S} $ & 145 & 138 & 66 & 65 & 115 & 77 & 86 & 349 &
298 & 259 \\
Exp & 141(1) & 144(1) & 46(3) & 47(4) &
117(2) & - & - & 630.5(0.6) & 393.92(0.24) &
335.3(0.1) \\
\hline
$ 2 \, {\rm S} $ & 81 & 81 & 40 & 39 & 67 & 42 & 35 & 135 & 130 &
127 \\
Exp & 57 & - & - & - &
92(5) & - & - & 165(103) & - & - \\
\hline
$ \Delta_{\rm avg} $ & 21 & 19 & 9 & 4 & 20 & - & 10 & 19  & 48 & 18 \\
\end{tabular}
\label{iperf}
\end{table}

\end{document}